\documentclass[12pt]{article}
\usepackage{amssymb}
\usepackage[dvipdfmx]{graphicx}
\usepackage[dvipdfmx]{color}

\usepackage{amsmath}
\usepackage{booktabs}
\usepackage{graphicx}
\usepackage{theorem}
\usepackage{times}
\usepackage{bm}
\usepackage{natbib}
\usepackage{amssymb}
\usepackage{color}
\usepackage[plain,noend]{algorithm2e}

\makeatletter
\renewcommand{\algocf@captiontext}[2]{#1\algocf@typo. \AlCapFnt{}#2} % text of caption
\def\@algocf@capt@plain{top}
\renewcommand{\algocf@makecaption}[2]{%
  \addtolength{\hsize}{\algomargin}%
  \sbox\@tempboxa{\algocf@captiontext{#1}{#2}}%
  \ifdim\wd\@tempboxa >\hsize%     % if caption is longer than a line
    \hskip .5\algomargin%
    \parbox[t]{\hsize}{\algocf@captiontext{#1}{#2}}% then caption is not centered
  \else%
    \global\@minipagefalse%
    \hbox to\hsize{\box\@tempboxa}% else caption is centered
  \fi%
  \addtolength{\hsize}{-\algomargin}%
}
\makeatother

\def\L{{{\cal L}}}

\def\col{{\bf col}}

\def\Var{{\rm Var}}

\def\ga{{\gamma}}
\def\de{{\delta}}
\def\ep{{\varepsilon}}

\def\bbe{{\text{\boldmath $\beta$}}}

\def\bep{{\text{\boldmath $\varepsilon$}}}

\def\bpsi{{\text{\boldmath $\psi$}}}

\def\psih{{\widehat \psi}}

\def\bbeh{{\widehat \bbe}}

\def\bpsih{{\widehat \bpsi}}

\def\Si{{\Sigma}}

\def\bSi{{\text{\boldmath $\Si$}}}

\def\a{{\text{\boldmath $a$}}}

\def\j{{\text{\boldmath $j$}}}

\def\u{{\text{\boldmath $u$}}}
\def\v{{\text{\boldmath $v$}}}

\def\x{{\text{\boldmath $x$}}}
\def\y{{\text{\boldmath $y$}}}

\def\A{{\text{\boldmath $A$}}}
\def\B{{\text{\boldmath $B$}}}
\def\C{{\text{\boldmath $C$}}}
\def\D{{\text{\boldmath $D$}}}
\def\E{{\text{\boldmath $E$}}}

\def\G{{\text{\boldmath $G$}}}
\def\H{{\text{\boldmath $H$}}}
\def\I{{\text{\boldmath $I$}}}
\def\J{{\text{\boldmath $J$}}}
\def\K{{\text{\boldmath $K$}}}
\def\L{{\text{\boldmath $L$}}}

\def\P{{\text{\boldmath $P$}}}
\def\Q{{\text{\boldmath $Q$}}}
\def\R{{\text{\boldmath $R$}}}

\def\V{{\text{\boldmath $V$}}}
\def\W{{\text{\boldmath $W$}}}
\def\X{{\text{\boldmath $X$}}}

\def\Z{{\text{\boldmath $Z$}}}

\def\xb{{\overline \x}}

\def\yo{{\overline y}}

\def\Re{{\mathbb{R}}}
\def\tr{{\rm tr\,}}
\def\diag{{\rm diag\,}}

\def\Cov{{\bf Cov\,}}

\def\[{{\text{\boldmath $[$}}}
\def\]{{\text{\boldmath $]$}}}

\def\zero{{\bf\text{\boldmath $0$}}}

\def\|{{\,|\,}}

\def\/{{\Bigr/\!\!}}

\def\1r{{\rm (1)}}
\def\2r{{\rm (2)}}
\def\3r{{\rm (3)}}
\def\4r{{\rm (4)}}
\def\5r{{\rm (5)}}

\def\non{{\nonumber}}
\def\bPt{{\widetilde \P}}
\def\col{{{\rm\bf col}}}
\def\mat{{{\rm\bf mat}}}
\def\bCt{{\widetilde \C}}
\def\bDt{{\widetilde \D}}
\def\bBt{{\widetilde \B}}
\def\bKt{{\widetilde \K}}

\newtheorem{thm}{Theorem}[section]
\newtheorem{lem}{Lemma}[section]

\newtheorem{prp}{Proposition}[section]
\theorembodyfont{\rmfamily}

\begin{document}
\title{General Unbiased Estimating Equations for Variance Components in Linear Mixed Models }

\author{
T. Kubokawa\thanks{Faculty of Economics, University of Tokyo, 7-3-1 Hongo, Bunkyo-ku, Tokyo 113-0033, { tatsuya@e.u-tokyo.ac.jp}},\ 
S. Sugasawa\footnote{Center for Spatial Information Science, The University of Tokyo, 5-1-5 Kashiwanoha, Kashiwa-shi, Chiba 277-8568,  {sugasawa@csis.u-tokyo.ac.jp}},\ 
H. Tamae\footnote{Nospare Inc., 2-7-13, Kita-aoyama, Minato-ku, Tokyo 107-0061, {taisho.1603@gmail.com}}\ \ and
\vspace{2mm}
S. Chaudhuri\footnote{Department of Statistics and Applied Probability, National University of Singapore, Block S16, Level 7, 6 Science Drive 2, 117546, SINGAPORE, {sanjay@stat.nus.edu.sg}}\\
{\it\normalsize   The University of Tokyo, Nospare Inc. and National University of Singapore}
}
\maketitle

\begin{abstract}
This paper introduces a general framework for estimating variance components in the linear mixed models via general unbiased estimating equations, which include some well-used estimators such as the restricted maximum likelihood estimator.
We derive the asymptotic covariance matrices and second-order biases under general estimating equations without assuming the normality of the underlying distributions and identify a class of second-order unbiased estimators of variance components.
It is also shown that the asymptotic covariance matrices and second-order biases do not depend on whether the regression coefficients are estimated by the generalized or ordinary least squares methods.
We carry out numerical studies to check the performance of the proposed method based on typical linear mixed models.

\par\vspace{4mm}
\end{abstract}

\section{Introduction}

Linear mixed models are widely used in a variety of scientific areas such as small area estimation (Rao and Molina, 2015), longitudinal data analysis (Verbeke and Molenberghs, 2006) and meta-analysis (Boreinstein et al., 2009), and estimation of variance components play an essential role in fitting the models.
Estimation of variance components has a long history, and various methods have been suggested in the literature. 
For example, the analysis of variance estimation (ANOVA), the minimum norm quadratic unbiased estimation (MINQUE), the maximum likelihood estimation (ML), and the restricted maximum likelihood estimation (REML) are well-known methods.
See Rao and Kleffe (1988) and Searle, Casella, and McCulloch (1992) for the details.

This paper is motivated by the derivation of the restricted maximum likelihood estimator.
In the linear mixed models, the maximum likelihood (ML) estimator of variance components $\psi$ is the solution of the likelihood equation $S(\psi, \bbeh)=0$ where $S(\cdot,\cdot)$ is the score equation and $\bbeh$ is the generalized least squares (GLS) estimator of regression coefficients $\bbe$.
Although $E\{S(\psi,\bbe)\}=0$ because $S(\psi,\bbe)$ is the score function, after substituting the estimator $\bbeh$ we have $E\{S(\psi, \bbeh)\}=h(\psi)$, which is not zero.
Despite $S(\psi, \bbeh)$ is asymptotically unbiased, the bias is not negligible under moderate sample sizes, which may lead to undesirable estimation performance. 
To overcome the issue, the corrected equation is $S(\psi, \bbeh)-h(\psi)=0$, and the solution of the equation gives the restricted maximum likelihood (REML) estimator, which is known to have better performance than the ML estimator. 
As noted later, the estimating equation is still valid without normality as long as some standard moment assumptions are met.

In this paper, we extend the idea of the unbiased estimating equations to more general situations, where $S(\cdot,\cdot)$ is not necessarily the score function and the underlying distribution is not necessarily normal.  
We suggest the general class of estimating equations for estimating parameters in covariance matrices of random effects and error terms without assuming the normality.
This class includes the restricted maximum likelihood estimator and the Fay-Herriot estimator (Fay and Herriot, 1979), and the Prasad-Rao estimator (Prasad and Rao, 1990), which have widely used in the small area estimation.
We first provide unified formulas of the asymptotic covariance matrices and second-order biases without assuming the normality.
The resulting important observation is that the asymptotic covariance matrices and second-order biases do not depend on whether the regression coefficients are estimated by the generalized or ordinary least squares methods, suggesting constructing a simpler estimating equation by using the ordinary least squares estimator.
Moreover, owing to the explicit formula for the second-order bias, we derive conditions to ensure that the resulting estimator is second-order unbiased without normality assumption.  
This is the main contribution of this work since the detailed derivation of the second-order asymptotic properties is quite tricky.  
We also apply the general theory to two important classes of linear mixed models, Fay-Herriot (Fay and Herriot, 1979) and nested error regression (Battese et al., 1988) models, and the numerical performance of the resulting estimators is investigated through simulation studies.

This paper is organized as follows:
The general unbiased estimating equations are introduced in Section \ref{sec:main} with the second-order biases and asymptotic covariance matrices of the resulting estimators.
Some specific estimators and their asymptotic properties are given in Section \ref{sec:est}.
A numerical investigation is given in Section \ref{sec:sim}, and all the proofs are given in the Appendix.

\section{General Estimating Equations for Variance Components}
\label{sec:main}

\subsection{Settings and restricted maximum likelihood estimator}
Consider the linear mixed model 
$$
\y=\X\bbe+\Z\v + \bep,
$$
where $\y$ is an observable $N$-variate vector, $\bbe$ is a $p$-variate vector of unknown regression coefficients, and $\X$ is an $N\times p$ known matrix of covariates, and $\Z$ is an $N\times m$ design matrix.
Here, $\v$ is a vector of random effects, and $\bep$ is a vector of sampling errors.
It is only assumed that $\v$ and $\bep$ are mutually independent and distributed as $E(\v)=\zero$, $\Cov(\v)=\R_v(\bpsi)=\R_v$, $E(\bep)=\zero$ and $\Cov(\bep)=\R_e(\bpsi)=\R_e$, where $\bpsi=(\psi_1, \ldots, \psi_k)^\top$ is a vector of unknown parameters including variance components.
Also, the fourth moments are described as $E[\{(\R_e^{-1/2}\bep)_i\}^4]=K_e+3$ and $E[\{(\R_v^{-1/2}\v)_i\}^4]=K_v+3$, where $(\a)_i$ is the $i$-th element of vector $\a$, and $\A^{1/2}$ is the symmetric root matrix of matrix $\A$.
Then, $E(\y)=\X\bbe$ and $\Cov(\y)=\bSi=\R_e(\bpsi)+\Z\R_v(\bpsi)\Z^\top$.
Throughout the paper, we use the simple notations $\partial_a =\partial/\partial \psi_a$, $\bSi_{(a)}=\partial_a\bSi$ and $\bSi_{(ab)}=\partial_a\partial_b\bSi$ for $a, b=1, \ldots, k$.

Under the normality, the maximum likelihood estimator of $\bpsi$ is the solution of the equations. 
$$
(\y-\X\bbeh)^\top\bSi^{-1}\bSi_{(a)}\bSi^{-1}(\y-\X\bbeh){ -\tr(\bSi^{-1}\bSi_{(a)})=0}, \ \ \ \ a=1, \ldots, k,
$$
where $\bbeh=(\X^\top\bSi^{-1}\X)^{-1}\X^\top\bSi^{-1}\y$ is the generalized least squares estimator.
The above estimating equations are approximately unbiased under large $N$, but the bias is not necessarily negligible under moderate $N$, leading to bias in the resulting estimator of $\bpsi$. 
As a solution, the restricted maximum likelihood (REML) estimator of $\bpsi$ has been widely used as the solution of the equations. 
$$
(\y-\X\bbeh)^\top\bSi^{-1}\bSi_{(a)}\bSi^{-1}(\y-\X\bbeh){ -\tr(\P \bSi_{(a)})=0}, \ \ \ \ a=1, \ldots, k,
$$
where $\P=\bSi^{-1} - \bSi^{-1}\X(\X^\top\bSi^{-1}\X)^{-1}\X^\top\bSi^{-1}$.
An essential property of the above estimating equations is that they are exactly unbiased. 
Moreover, a key observation for the equations is that the unbiasedness property is still valid without the normality, and it only requires the moment assumptions.
In this work, we generalize the REML method for estimating the variance components $\bpsi$, that is, we consider the general class of unbiased estimating equations and develop a unified asymptotic theory for the resulting estimator of $\bpsi$.

\subsection{General estimating equations for variance parameters}
Let $\bbeh=\L\y$ be a linear unbiased estimator of $\bbe$, where $\L=\L(\bpsi)$ is a $p\times N$ matrix of functions of $\bpsi$ and satisfies $\L\X=\I$.
Let $\W_a=\W_a(\bpsi)$ be an $N\times N$ matrix of functions of $\bpsi$ for $a=1, \ldots, k$.
The expectation $E\{(\y-\X\bbeh)^\top\W_a(\y-\X\bbeh)\}$ is $\tr(\Q^\top\W_a\Q\bSi)$ for $\Q=\I-\X\L$, which gives the general estimating equations
\begin{equation}
\y^\top\Q^\top\W_a\Q\y - \tr(\Q^\top\W_a\Q\bSi) = 0,
\quad a=1, \ldots, k.
\label{eqn:ee}
\end{equation}
For example, the choice of $\W_a=\bSi^{-1}\bSi_{(a)}\bSi^{-1}$ leads to the REML estimation, and other choices of $\W_a$ leads to different estimator of $\bpsi$. 
In the following theorem, we provide the second-order bias and asymptotic covariance matrix of the general estimator $\bpsih$ as the solution of (\ref{eqn:ee}).
For notational convenience, we define $\W_{a(b)}=\partial_b\W_a$ and $\W_{a(bc)}=\partial_b\partial_c \W_a$ for $a, b, c =1, \ldots, k$.
Also, $(\V)_{ab}$ and $(\V)^{ab}$ denote the $(a,b)$-th element of matrix $\V$ and the inverse $\V^{-1}$.
Define $k\times k$ matrices $\A$ and $\B$ by
\begin{equation}
\begin{split}
(\A)_{ab} =& \tr(\W_a\bSi_{(b)}),\quad
(\B)_{ab}=\tr(\W_a\bSi\W_b\bSi), \\
(\bBt)_{ab}=& K_eh_e( \W_a,\W_b)+K_vh_v( \W_a,\W_b),
\end{split}
\label{eqn:AB}
\end{equation}
where for matrices $\C$ and $\D$,
\begin{align*}
h_e( \C, \D)=&\sum_{i=1}^N (\R_e^{1/2}\C\R_e^{1/2})_{ii} \cdot(\R_e^{1/2}\D\R_e^{1/2})_{ii},\\
h_v( \C, \D)=&\sum_{i=1}^m (\R_v^{1/2}\Z^\top\C\Z\R_v^{1/2})_{ii}\cdot (\R_v^{1/2}\Z^\top\D\Z\R_v^{1/2})_{ii}.
\end{align*}

\begin{thm}
\label{thm:1}
Assume that $(\L\X^\top\W_a\X\L)_{ij}=O(N^{-1})$, $\L\bSi\L^\top=O(N^{-1})$, and $(\X\L)_{ij}=O(N^{-1})$ as $N\to\infty$.
Then, $\Cov(\bpsih)=2\A^{-1}\B\A^{-1}+\A^{-1}\bBt\A^{-1}+O(N^{-3/2})$ and 
\begin{align}
E(\bpsih-\bpsi)
=& 2\A^{-1}\col_a(\K_a\A^{-1}-\H_a\A^{-1}\B\A^{-1})\non\\
&+\A^{-1}\col_a(\bKt_a\A^{-1}-\H_a\A^{-1}\bBt\A^{-1})+O(N^{-3/2}),
\label{eqn:bias}
\end{align}
where $\col_a(x_a)=(x_1, \ldots, x_k)^\top$ is a column vector with the $a$-th element $x_a$, and $(\K_a)_{bc}=\tr(\W_{a(b)}\bSi\W_c\bSi)$, $(\H_a)_{bc}=\tr(\W_{a(b)}\bSi_{(c)})+2^{-1}\tr(\W_a\bSi_{(bc)})$ and $(\bKt_a)_{bc}=K_eh_e(\W_{a(b)},\W_c)+K_vh_v(\W_{a(b)},\W_c)$.
\end{thm}

Two typical choices of $\L$ are $\L^{\rm G}=(\X^\top\bSi^{-1}\X)^{-1}\X^\top\bSi^{-1}$ and $\L^{\rm O}=(\X^\top\X)^{-1}\X^\top$, which give the generalized least squares (GLS) and ordinary least squares (OLS) estimators of $\bbe$, respectively, denoted by $\bbeh^{\rm G}$ and $\bbeh^{\rm O}$.
However, Theorem \ref{thm:1} tells us that the second-order bias and the asymptotic covariance matrix do not depend on such a choice of $\L$.
This is an essential observation from Theorem \ref{thm:1}, and the specific form of $\bbeh$ in the estimating equation (\ref{eqn:ee}) is irrelevant to the asymptotic properties of $\bpsih$ as long as $\bbeh$ is unbiased.
Hence, it would be better to use a simpler form of $\bbeh$, so in what follows, we employ $\L=(\X^\top\X)^{-1}\X^\top$, corresponding to the ordinary least squares estimators of $\bbe$.
On the other hand, the choice of $\W_{a}$ affects the asymptotic properties.
 
The second-order unbiasedness is one of the desirable properties of estimators $\bpsih$.
From Theorem \ref{thm:1}, we need to use $\W_a$ such that the leading term in (\ref{eqn:bias}) is $0$ to achieve second-order unbiasedness of $\bpsih$. 
In typical linear mixed models such as the Fay-Herriot (Fay and Herriot, 1979) and nested error regression (Battese et al., 1988) models, the covariance matrix $\bSi$ is a linear function of $\bpsi$.
In this case, $\bSi_{(bc)}=0$, which simplifies the condition for the second-order unbiasedness in (\ref{eqn:bias}), because $(\H_a)_{bc}=\tr(\W_{a(b)}\bSi_{(c)})$.
When $K_e=K_v=0$, the estimator $\bpsih$ is second-order unbiased if 
\begin{equation}
\K_a=\H_a\A^{-1}\B.
\label{eqn:cond}
\end{equation}
This condition is investigated in the next section for some specific choices of $\W_a$.

\section{Specific Estimators and Their Asymptotic Properties}
\label{sec:est}

\subsection{Three estimators}

We now describe some specific estimators of $\bpsi$ and provide their asymptotic variances and biases.
In what follows, we assume that $\bSi$ is a linear function of $\bpsi$, which are satisfied in typical linear mixed models such as the Fay-Herriot and nested error regression models.
We here consider the three candidates for $\W_a$; $\W_a^{\rm RE}=\bSi^{-1}\bSi_{(a)}\bSi^{-1}$, $\W_a^{\rm FH}=(\bSi^{-1}\bSi_{(a)}+ \bSi_{(a)}\bSi^{-1})/2$ and $\W_a^{\rm Q}=\bSi_{(a)}$, which are motivated from the REML estimator, the Fay-Herriot moment estimator (Fay and Herriot, 1979) and the Prasad-Rao unbiased estimator (Prasad and Rao, 1990) under a simple linear mixed models called Fay-Herriot model. 
The estimators induced from $\W_a^{\rm RE}$, $\W_a^{\rm FH}$ and $\W_a^{\rm Q}$ are called here the REML-type, FH-type and PR-type estimators, respectively.
From Theorem \ref{thm:1}, we can derive the asymptotic properties of the three estimators.
When $\bSi$ is a linear function of $\bpsi$, the asymptotic variances and second-order biases are simplified in the case of $K_e=K_v=0$, which is satisfied in the normal distributions.

\begin{prp}\label{prp:bias}
Assume the conditions in Theorem \ref{thm:1} and that $\bSi$ is a linear function of $\bpsi$.
Also assume that $K_e=K_v=0$.
Let $\bpsih^{\rm RE}$, $\bpsih^{\rm FH}$ and $\bpsih^{\rm Q}$ be the estimators based on $\W_a^{\rm RE}$, $\W_a^{\rm FH}$ and $\W_a^{\rm Q}$, respectively.
Then the following results hold.

{\rm (a)}\ 
REML-type estimator $\bpsih^{\rm RE}$ is second-order unbiased and has the asymptotic covariance matrix $2 \A _{\rm RE}^{-1}$, where $( \A _{\rm RE})_{ij}=\tr(\bSi^{-1}\bSi_{(i)}\bSi^{-1}\bSi_{(j)})$.

{\rm (b)}\ 
FH-type estimator $\bpsih^{\rm FH}$ is not second-order unbiased.
The asymptotic covariance matrix is $2 \A _{\rm FH}^{-1} \B _{\rm FH} \A _{\rm FH}^{-1}$ for $( \A _{\rm FH})_{ij}=\tr(\bSi ^{-1}\bSi_{(i)}\bSi_{(j)})$ and $( \B _{\rm FH})_{ij}=\{\tr(\bSi_{(i)}\bSi_{(j)})+\tr(\bSi^{-1}\bSi_{(i)}\bSi \bSi_{(j)})\}/2$.
The second-order bias is $2\A _{\rm FH}^{-1}\col_a(\K_a\A _{\rm FH}^{-1}-\H_a\A _{\rm FH}^{-1}\B _{\rm FH}\A _{\rm FH}^{-1})$ for $(\K_a)_{bc}=-\tr\{\bSi_{(a)}(\bSi_{b)}+\bSi\bSi_{(b)}\bSi^{-1})\bSi_{(c)}\bSi^{-1}\}/2$ and $(\H_a)_{bc}=-\tr\{\bSi_{a)}\bSi_{(b)}\bSi^{-1})\bSi_{(c)}\bSi^{-1}$.

{\rm (c)}\ 
PR-type estimator $\bpsih^{\rm Q}$ is second-order unbiased and has the asymptotic covariance matrix $2 \A _{\rm Q}^{-1} \B _{\rm Q} \A _{\rm Q}^{-1}$, where $( \A _{\rm Q})_{ij} = \tr(\bSi_{(i)}\bSi_{(j)})$ and $( \B _{\rm Q})_{ij}=\tr(\bSi_{(i)}\bSi\bSi_{(j)}\bSi )$.
\end{prp}

In Proposition \ref{prp:bias}, the linearity of $\bSi(\bpsi)$ on $\bpsi$ is only used to compute the second-order bias.  
The expressions for the asymptotic covariances hold, in general, without such constraints.  
Without assuming $K_e=K_v=0$, the estimator $\bpsih^{\rm RE}$ has the second-order bias, while $\bpsih^{\rm Q}$ remains second-order unbiased.

It is noted that the REML-type is the most efficient in the normal distributions, which corresponds to the case of $K_e=K_v=0$.
This implies that the following inequality holds for any $\W_a$:
\begin{align}
[\mat_{a,b}\{\tr(\W_a\bSi_{(b)})\}]^{-1} &\mat_{a,b}\{\tr(\W_a\bSi\W_b\bSi)\} [\mat_{a,b}\{\tr(\W_a\bSi_{(b)})\}]^{-1}\non\\
&\geq [\mat_{a,b}\{\tr(\bSi^{-1}\bSi_{(a)}\bSi^{-1}\bSi_{(b)})\}]^{-1},
\label{eqn:ef}
\end{align}
where $\mat_{ab}\{x_{ab}\}$ is a $k\times k$ matrix with the $(a,b)$-th element $x_{ab}$.
However, it should be remarked that REML is not necessarily efficient without assuming $K_e=0$ ans $K_v=0$.

\subsection{Detailed properties under two useful models}
We provide more detailed formulas for the asymptotic covariances (or variances) and second-order biases under two models, Fay-Herriot and nested error regression models.

The first model is the Fay-Herriot model, which corresponds to $\y=(y_1, \ldots, y_m)^\top$, $N=m$, $\R_v=\psi_1\I_m$, $\R_e=\D=\diag(D_1, \ldots, D_m)$ and $\bSi=\psi_1\I_m+\D$ for known $D_i$'s.
The following proposition can be derived from Theorem \ref{thm:1}:

\begin{prp}
\label{prp:FH}
In the Fay-Herriot model, estimator $\psih_1$ is the solution of $(\ref{eqn:ee})$ for diagonal matrix $\W_1$. 
Without assuming $K_e=K_v=0$, the asymptotic variance of $\psih_1$ is 
$$
\Var(\psih_1)\approx2{\tr(\W_1\bSi\W_1\bSi)\over \{\tr(\W_1)\}^2}+{K_e \tr(\W_1^2\D^2)+\psi_1^2K_v\tr(\W_1^2) \over \{\tr(\W_1)\}^2},
$$ 
and the second-order bias is 
\begin{align*}
{\rm Bias}(\psih_1)
\approx &2{\tr(\W_{1(1)}\bSi\W_1\bSi)-\tr(\W_{1(1)})\tr(\W_1\bSi\W_1\bSi)\over \{\tr(\W_1)\}^2} \\
&+{K_e \tr(\W_{1(1)}\W_1\D^2)+\psi_1^2K_v\tr(\W_{1(1)}\W_1) \over \{\tr(\W_1)\}^2}\\
&-{\tr(\W_{1(1)})\{K_e \tr(\W_1^2\D^2)+\psi_1^2K_v\tr(\W_1^2) \}\over \{\tr(\W_1)\}^3}.
\end{align*}
\end{prp}

In this model, the inequality (\ref{eqn:ef}) is expressed as
$$
{\tr(\W_1\bSi\W_1\bSi)\over \{\tr(\W_1)\}^2} \geq {1\over \tr(\bSi^{-2})},
$$
or $\tr(\W_1\bSi\W_1\bSi)\tr(\bSi^{-2})\geq \{\tr(\W_1)\}^2$.
This inequality can be directly proved by using the Cauchy-Schwarz inequality.

The second example is the nested error regression model.
Let $\Z=\text{block diag}(\j_{n_1}, \ldots, \j_{n_m})$ for $\j_{n_i}=(1, \ldots, 1)^\top\in \Re^{n_i}$, and let $\G=\text{block diag}(\J_{n_1}, \ldots, \J_{n_m})$ for $\J_{n_i}=\j_{n_i}\j_{n_i}^\top$.
This model corresponds to $N=\sum_{i=1}^m n_i$, $\R_v=\psi_1\G$, $\R_e=\psi_2\I_N$ and $\bSi=\psi_1\G+\psi_2\I_N$.
Note that $\bSi_{(1)}=\G$ and $\bSi_{(2)}=\I_N$.
Then,
$$
\A=\begin{pmatrix}\tr(\W_1\G) & \tr(\W_1)\\ \tr(\W_2\G) & \tr(\W_2)\end{pmatrix},\ 
\B= \begin{pmatrix}\tr(\W_1\bSi\W_1\bSi) & \tr(\W_1\bSi\W_2\bSi)\\ \tr(\W_1\bSi\W_2\bSi) & \tr(\W_2\bSi\W_2\bSi)\end{pmatrix},
$$
where $(\bBt)_{ab}=\psi_2^2 K_e\sum_{i=1}^N (\W_a)_{ii} (\W_b)_{ii}+\psi_1^2 K_v\sum_{i=1}^m (\Z^\top\W_a\Z)_{ii} (\Z^\top\W_b\Z)_{ii}$.
The following proposition is provided from Theorem \ref{thm:1}:

\begin{prp}
\label{prp:NER}
In the nested error regression model, estimator $\bpsih$ is the solution of $(\ref{eqn:ee})$. 
Without assuming $K_e=K_v=0$, the asymptotic covariance matrix of $\bpsih$ is 
\begin{equation}
\Cov(\bpsih)\approx 2\A^{-1}\B\A^{-1}+ \A^{-1}\bBt\A^{-1},
\label{eqn:NERcov}
\end{equation}
and the second-order bias of $\bpsih$ is
\begin{align}
&{\bf Bias}(\bpsih)\label{eqn:NERbias}
\\
&\approx 
2\A^{-1}\begin{pmatrix}\tr(\K_1\A^{-1})-\tr(\H_1\A^{-1}\B\A^{-1})\\
\tr(\K_2\A^{-1})-\tr(\H_2\A^{-1}\B\A^{-1})\end{pmatrix}
+\A^{-1} \begin{pmatrix} \tr(\bKt_1\A^{-1})-\tr(\H_1\A^{-1}\bBt\A^{-1})\\
\tr(\bKt_2\A^{-1})-\tr(\H_2\A^{-1}\bBt\A^{-1})\end{pmatrix},\non
\end{align}
where $(\K_a)_{bc}=\tr(\W_{a(b)}\bSi\W_c\bSi)$, $(\H_a)_{bc}=\tr(\W_{a(b)}\bSi_{(c)})$ and \\
$(\bKt_a)_{bc}=\psi_2^2K_e\sum_{i=1}^N (\W_{a(b)})_{ii} (\W_c)_{ii}+ \psi_1^2K_v\sum_{i=1}^m (\Z^\top\W_{a(b)}\Z)_{ii} (\Z^\top\W_c\Z)_{ii}$ for $a=1, 2$. 
\end{prp}

% simulation 
\section{Simulation studies}
\label{sec:sim}

We here investigate the finite-sample performance of estimators obtained from the estimating equation (\ref{eqn:ee}).
We consider the mean squared errors (MSE) of the four estimators: the REML estimator $\bpsih^{\rm RE}$, the REML estimator with ordinary least squares estimator of $\bbe$ in (\ref{eqn:ee}) denoted by $\bpsih^{\rm ORE}$, the FH-type estimator $\bpsih^{\rm FH}$ and the PR-type estimator $\bpsih^{\rm Q}$, which are denoted by RE, ORE, FH and Q in Tables \ref{table:FHsim} and \ref{table:NERMsim}.
In what follows, we considered two typical linear mixed models, the Fay-Herriot model and the nested error regression model.
The detailed forms of estimators are provided in the Appendix.

We first consider the Fay-Herriot model studied in Fay and Herriot (1979) as a simple area-level linear mixed model, where the model is described above Proposition \ref{prp:FH}.
In this model, we have $\bSi =\psi_1 \I_m +\diag(D_1, \ldots,D_m)$ with known $D_i$'s, and we use the same setup as in Datta, Rao and Smith (2005) in the normal distributions. 
We adopt the $D_i$-pattern $(0.7, 0.6, 0.5, 0.4, 0.3)$ in Datta, et al. (2005).
Note that $\psih_1^{\rm Q}$ is identical to the Prasad-Rao estimator suggested by Prasad and Rao (1990), a widely used unbiased estimator of $\psi$ in this context, in this model.
We computed the square-roots of MSEs for the four estimators based on 10,000 replications and reported the results in Table \ref{table:FHsim} for $\psi_1=0.7, 1.0, 3.0$.
It is observed that $\psih_1^{\rm RE}$ and $\psih_1^{\rm ORE}$ perform in the same way, which supports the fact given in Theorem \ref{thm:1}.
Moreover, $\psih_1^{\rm FH}$ has a similar performance to the two restricted maximum likelihood estimator $\psih_1^{\rm RE}$ and $\psih_1^{\rm ORE}$, whereas $\psih_1^{\rm Q}$ performs the worst among the four estimators.

% Table 1
\begin{table}[htbp]
\caption{Square-root of MSE for the four estimators in the Fay-Herriot model}
\begin{center}
$
{\renewcommand\arraystretch{1.1}\small
\begin{array}{c@{\hspace{2mm}}
       c@{\hspace{2mm}}
       c@{\hspace{2mm}}
       c@{\hspace{2mm}}
       c@{\hspace{2mm}}
       c@{\hspace{2mm}}
       r
       }
\hline
\text{$\psi_1$} &&\text{RE}&\text{ORE}&\text{FH}&\text{Q}\\
\hline
0.7&& 0.3044& 0.3044& 0.3062& 0.3125\\
1.0&& 0.3873& 0.3873& 0.3881& 0.3927\\
3.0&& 0.9276& 0.9276& 0.9283& 0.9306\\
\hline
\end{array}
}
$
\end{center}
\label{table:FHsim}
\end{table}

We next consider the nested error regression model studied in Battese, Harter and Fuller (1988) as a unit-level random intercept model: $y_{ij}=x_{ij}^\top\beta+v_i + \ep_{ij}$ for $i=1, \ldots, m$ and $j=1, \ldots,n_i$ with $E(v_i)=0$, $\Var(v_i)=\psi_1$, $E(\ep_{ij})=0$, $\Var(\ep_{ij})=\psi_2$ and $p=3$, where $m=15$ clusters are equally divided into five groups, and the sample sizes $n_i$ are the same for clusters within the same group.
We set the $n_i$-pattern to $(5,5,6,6,7)$, so that the total sample size is $N=87$.
Since the original Prasad-Rao estimator $\psih_1^{\rm PR}$ is different from $\psih_1^{\rm Q}$, we include it for comparison.
In the same way as the previous study, we computed the square roots of MSEs for the five estimators based on 5,000 replications. We reported the results in Table \ref{table:NERMsim} for $\psi_1=0.5, 1.0, 2.0$ and $\psi_2=1$ under normality.
Again, we can observe that $\psih_1^{\rm RE}$ and $\psih_1^{\rm ORE}$ perform almost in the same way.
However, we found that the solution $\psih_1^{\rm RE}$ did not exist in some replications.
In contrast, solutions $\psih_1^{\rm FH}$ and $\psih_1^{\rm ORE}$ existed in all the replications.
We also observed that $\psih_1^{\rm Q}$ did not perform well compared with the other methods.

\begin{table}[htbp]
\caption{Square-root of MSE for the five estimators in the nested error regression model}
\begin{center}
$
{\renewcommand\arraystretch{1.1}\small
\begin{array}{c@{\hspace{2mm}}
       r@{\hspace{2mm}}
       r@{\hspace{2mm}}
       r@{\hspace{2mm}}
       r@{\hspace{2mm}}
       r@{\hspace{2mm}}
       r@{\hspace{2mm}}
       r
       }
\hline
\text{$\psi_1$} &&\text{RE}&\text{ORE}&\text{FH}&\text{Q}&\text{PR}\\
\hline
0.5&& 0.2906& 0.2910& 0.2954& 0.3028& 0.2943\\
1.0&& 0.6149& 0.6164& 0.6384& 0.6582& 0.6349\\
2.0&& 1.2018& 1.2059& 1.2264& 1.2592& 1.2220\\
\hline
\end{array}
}
$
\end{center}
\label{table:NERMsim}
\end{table}

\section*{Acknowledgments}
%We would like to thank the Associate Editor and the two reviewers for many valuable comments and helpful suggestions, which led to an improved version of this paper.
This research was supported in part by Grant-in-Aid for Scientific Research (18K11188) from the Japan Society for the Promotion of Science.

\appendix
\section{Proofs}

\subsection{A preliminary lemma}

For the proof, we use the following lemma:

\begin{lem}
\label{lem:1}
Let $\u=\bep+\Z\v$.
Then, for matrices $\C$ and $\D$, it holds that 
\begin{equation}
E[\u^\top\C\u\u^\top\D\u]=2\tr(\C\bSi\D\bSi)+\tr(\C\bSi)\tr(\D\bSi) + K_e h_e(\C,\D) +K_v h_v(\C, \D),
\label{eqn:L1}
\end{equation}
for $h_e(\C,\D)$ and $h_v(\C, \D)$ defined above Theorem \ref{thm:1}.
\end{lem}

{\bf Proof}.\ \ It is demonstrated that $E[\u^\top\C\u\u^\top\D\u]=E[\bep^\top\C\bep\bep^\top\D\bep]+E[\v^\top\Z^\top\C\Z\v\v^\top\Z^\top\D\Z\v]+\tr(\C\R_e)\tr(\D\Z\R_v\Z^\top)+\tr(\D\R_e)\tr(\C\Z\R_v\Z^\top)+4\tr(\C\R_e\D\Z\R_v\Z^\top)$.
Let $\x=(x_1, \ldots, x_N)^\top=\R_e^{-1/2}\bep$, $\bCt=\R_e^{1/2}\C\R_e^{1/2}$ and $\bDt=\R_e^{1/2}\D\R_e^{1/2}$.
Then, $E[\x]=\zero$, $E[\x\x^\top]=\I_N$, $E[x_a^4]=K_e+3$, $a=1, \ldots, N$, and $E[\bep^\top\C\bep\bep^\top\D\bep]=E[\x^\top\bCt\x \x^\top\bDt\x]$.
Let $\de_{a=b=c=d}=1$ for $a=b=c=d$, and otherwise, $\de_{a=b=c=d}=0$.
The notation $\de_{a=b\not= c=d}$ is definced similarly.
It is observed that for $a, b, c, d =1, \ldots, N$, 
\begin{align*}
E[x_a&(\bCt)_{ab}x_b x_c(\bDt)_{cd}x_d]\\
=&
E[ x_a^4 (\bCt)_{aa}(\bDt)_{aa}\de_{a=b=c=d}+x_a^2x_c^2 (\bCt)_{aa}(\bDt)_{cc}\de_{a=b\not= c=d} + 2 x_a^2x_b^2(\bCt)_{ab}(\bDt)_{ab}\de_{a=c\not= b=d}]
\\
=& (K_e+3) (\bCt)_{aa}(\bDt)_{aa}\de_{a=b=c=d}+(\bCt)_{aa}(\bDt)_{cc}\de_{a=b\not= c=d} + 2 (\bCt)_{ab}(\bDt)_{ab}\de_{a=c\not= b=d}
\\
=& K_e(\bCt)_{aa}(\bDt)_{aa}\de_{a=b=c=d}+(\bCt)_{aa}(\bDt)_{cc}\de_{a=b}\de_{c=d} + 2 (\bCt)_{ab}(\bDt)_{ab}\de_{a=c}\de_{b=d},
\end{align*}
which implies that
$$
\sum_{a, b, c, d}E[x_a(\bCt)_{ab}x_b x_c(\bDt)_{cd}x_d]
= K_e \sum_{a=1}^N (\bCt)_{aa}(\bDt)_{aa}+ \sum_{a=1}^N(\bCt)_{aa}\sum_{c=1}^N(\bDt)_{cc} + 2 \sum_{a=1}^N\sum_{b=1}^N(\bCt)_{ab}(\bDt)_{ab},
$$
or 
$$
E[\bep^\top\C\bep\bep^\top\D\bep]=2\tr(\C\R_e\D\R_e)+\tr(\C\R_e)\tr(\D\R_e)
+K_e h_e(\C, \D).
$$
Similarly, 
\begin{align*}
E[\v^\top&\Z^\top\C\Z\v\v^\top\Z^\top\D\Z\v]\\
=&2\tr(\C\Z\R_v\Z^\top\D\Z\R_v\Z^\top)+\tr(\C\Z\R_v\Z^\top)\tr(\D\Z\R_v\Z^\top) + K_v h_v(\C,\D).
\end{align*}
Thus, we have
\begin{align*}
E[\u^\top\C\u\u^\top\D\u]=&2\tr(\C\R_e\D\R_e)+\tr(\C\R_e)\tr(\D\R_e)+2\tr(\C\Z\R_v\Z^\top\D\Z\R_v\Z^\top)
\\
&+\tr(\C\Z\R_v\Z^\top)\tr(\D\Z\R_v\Z^\top) +\tr(\C\R_e)\tr(\D\Z\R_v\Z^\top)\\
&+\tr(\D\R_e)\tr(\C\Z\R_v\Z^\top)+4\tr(\C\R_e\D\Z\R_v\Z^\top)\\
&+K_e h_e(\C, \D)+ K_v h_v(\C,\D),
\end{align*}
which can be rewritten as the expression in (\ref{eqn:A1}) for $\bSi=\R_e+\Z\R_v\Z^\top$.
\hfill$\Box$

\subsection{Proof of Theorem \ref{thm:1}}

For $a=1, \ldots, k$, let $\ell_a=\y^\top\C_a\y - \tr(\D_a)$ for $\C_a=\Q^\top\W_a\Q$ and $\D_a=\Q^\top\W_a\Q\bSi$.
For $\u=\y-\X\bbe=\bep+\Z\v$, $\ell_a$ is rewritten as $\ell_a=\u^\top\C_a\u-\tr(\D_a)$. 
By the Taylor series expansion, 
$$
0=\col_a(\ell_a) + \mat_{ab}(\ell_{a(b)})(\bpsih-\bpsi)
+{1\over 2}\col_a\Big\{\sum_{b=1}^k\sum_{c=1}^k \ell_{a(bc)} (\psih_b-\psi_b)(\psih_c-\psi_c)\Big\}
+O_p(N^{-1/2}),
$$
where $\mat_{ab}(x_{ab})$ is a $k\times k$ matrix with the $(a,b)$-th element $x_{ab}$.
Then,
$$
\bpsih-\bpsi
=- \{\mat_{ab}(\ell_{a(b)}) \}^{-1} \Big[ \col_a(\ell_a) 
+{1\over 2}\col_a\Big\{\sum_{b=1}^k\sum_{c=1}^k \ell_{a(bc)}(\psih_b-\psi_b)(\psih_c-\psi_c)\Big\}\Big] 
+O_p(N^{-3/2}).
$$
Since $\tr(\bSi\C_a)=\tr(\D_a)$, we have $\ell_a=\tr\{\C_a(\u\u^\top-\bSi)\}$. 
Also, $\ell_{a(b)}= \tr(\bSi\C_{a(b)}-\D_{a(b)})+\tr\{\C_{a(b)}(\u\u^\top-\bSi)\}$ and $\ell_{a(bc)}= \tr(\bSi\C_{a(bc)}-\D_{a(bc)})+\tr\{\C_{a(bc)}(\u\u^\top-\bSi)\}$.
Let $\A_1=\mat_{ab}\{\tr(\bSi\C_{a(b)}-\D_{a(b)})\}$ and $\A_0=\mat_{ab}[\tr\{\C_{a(b)}(\u\u^\top-\bSi)\}]$. 
It is noted that $\A_1=O(N)$, $\A_0=O_p(N^{1/2})$, $\tr(\bSi\C_{a(bc)}-\D_{a(bc)})=O(N)$ and $\tr\{\C_{a(bc)}(\u\u^\top-\bSi)\}=O_p(N^{1/2})$.
Then it can be seen that
$$
\{\mat_{ab}(\ell_{a(b)}) \}^{-1} = (\A_1+\A_0)^{-1}=\A_1^{-1}-\A_1^{-1}\A_0\A_1^{-1}+O_p(N^{-2}),
$$
so that
\begin{align*}
\bpsih-\bpsi =&
-\A_1^{-1} \col_a[\tr\{\C_a(\u\u^\top-\bSi)\}] + \A_1^{-1}\A_0\A_1^{-1} \col_a[\tr\{\C_a(\u\u^\top-\bSi)\}] 
\non\\
&-{1\over 2}\A_1^{-1}\col_a\Big\{\sum_{b=1}^k\sum_{c=1}^k \tr(\bSi\C_{a(bc)}-\D_{a(bc)})(\psih_b-\psi_b)(\psih_c-\psi_c)\Big\} 
+O_p(N^{-3/2}).
\end{align*}
It is noted that $(\C_a)_{ij}=(\Q^\top\W_a\Q)_{ij}=(\W_a)_{ij}+O(N^{-1})$, $(\C_{a(b)})_{ij}=(\W_{a(b)})_{ij}+O(N^{-1})$ and $(\C_{a(bc)})_{ij}=(\W_{a(bc)})_{ij}+O(N^{-1})$.
Then, $\tr(\C_a\bSi)=\tr(\W_a\bSi)+O(1)$, $\tr(\C_{a(b)}\bSi)=\tr(\W_{a(b)}\bSi)+O(1)$ and $\tr(\C_{a(bc)}\bSi)=\tr(\W_{a(bc)}\bSi)+O(1)$.
Since $\D_a=\C_a\bSi$, $\D_{a(b)}=\C_{a(b)}\bSi+\C_a\bSi_{(b)}$ and $\D_{a(bc)}=\C_{a(bc)}\bSi+\C_{a(b)}\bSi_{(c)}+\C_{a(c)}\bSi_{(b)}+\C_a\bSi_{(bc)}$, it is seen that $\tr(\D_{a(b)})=\tr(\W_{a(b)}\bSi)+\tr(\W_a\bSi_{(b)}) +O(1)$ and $\tr(\D_{a(bc)})=\tr(\W_{a(bc)}\bSi)+\tr(\W_{a(b)}\bSi_{(c)})+\tr(\W_{a(c)}\bSi_{(b)})+\tr(\W_a\bSi_{(bc)})+O(1)$.
Thus,
\begin{equation}
\begin{split}
\tr(\bSi\C_{a(b)}-\D_{a(b)})=& - \tr(\W_a\bSi_{(b)})+O(1),\\
\tr(\bSi\C_{a(bc)}-\D_{a(bc)})=& -\tr(\W_{a(b)}\bSi_{(c)})-\tr(\W_{a(c)}\bSi_{(b)})-\tr(\W_a\bSi_{(bc)}) +O(1).
\end{split}
\label{eqn:A1}
\end{equation}
Letting $\A=\mat_{ab}\{\tr(\W_a\bSi_{(b)})\}$, we have $\A_1=-\A+O(1)$.
Using Lemma \ref{lem:1}, we can approximate the covariance matrix of $\bpsih$ as
\begin{align*}
E[(\bpsih-\bpsi)(\bpsih-\bpsi)^\top]=& \A_1^{-1} \mat_{ab}( E[\tr\{\C_a(\u\u^\top-\bSi)\}\tr\{\C_b(\u\u^\top-\bSi)\}] ) \A_1^{-1} + O(N^{-3/2})\\
=& 2\A^{-1}\B\A^{-1}+\A^{-1}\bBt\A^{-1} + O(N^{-3/2}),
\end{align*} 
for $\B=\mat_{ab}\{\tr(\W_a\bSi\W_b\bSi)\}$ and $\bBt=\mat_{ab}\{K_eh_e(\W_a,\W_b)+K_vh_v(\W_a,\W_b)\}$.

The bias of $\bpsih$ is 
\begin{align*}
E(\bpsih-\bpsi) =&
-{1\over 2}\A^{-1}\col_a\Big[\sum_{b=1}^k\sum_{c=1}^k \{2\tr(\W_{a(b)}\bSi_{(c)})+\tr(\W_a\bSi_{(bc)})\} (\A^{-1}(2 \B+\bBt)\A^{-1})_{bc}\Big] \\
&+ E(\A^{-1}\A_0\A^{-1} \col_a[\tr\{\C_a(\u\u^\top-\bSi)\}] )
+O(N^{-3/2}).
\end{align*}
Concerning the second term in RHS, the $a$-th element of $E\{ (\A_0\A^{-1} \col_c[\tr\{\C_c(\u\u^\top-\bSi)\}])\}$ is
\begin{align*}
E\{ &(\A_0\A^{-1} \col_c[\tr\{\C_c(\u\u^\top-\bSi)\}])_a\}\\
=&\sum_{b=1}^k\sum_{c=1}^k E[\tr\{\C_{a(b)}(\u\u^\top-\bSi)\}(\A)^{bc}\tr\{\C_c(\u\u^\top-\bSi)\}]
\\
=&
\sum_{b=1}^k\sum_{c=1}^k \{2\tr(\W_{a(b)}\bSi\W_c\bSi) +K_eh_e(\W_{a(b)},\W_c)+K_vh_v(\W_{a(b)},\W_c)\}(\A)^{bc} + O(N^{-1}).
\end{align*}
Then,
\begin{align*}
E&(\bpsih-\bpsi)\non\\
=&\A^{-1}\col_a\Big(\sum_{b=1}^k\sum_{c=1}^k \{2\tr(\W_{a(b)}\bSi\W_c\bSi) +K_eh_e(\W_{a(b)},\W_c)+K_vh_v(\W_{a(b)},\W_c)\}(\A)^{bc}\Big)\non\\
& -{1\over 2}\A^{-1}\col_a\Big[\sum_{b=1}^k\sum_{c=1}^k \{2\tr(\W_{a(b)}\bSi_{(c)})+\tr(\W_a\bSi_{(bc)})\} (\A^{-1} (2 \B+\bBt)\A^{-1})_{bc}\Big] +O(N^{-3/2}),
\end{align*}
which provides the expression in (\ref{eqn:bias}) in Theorem \ref{thm:1}.

\subsection{Proof of Proposition \ref{prp:bias}}

Case of $\W_a=\bSi^{-1}\bSi_{(a)}\bSi^{-1}$. 
We have $\W_{a(b)}=-\bSi^{-1}\bSi_{(b)}\bSi^{-1}\bSi_{(a)}\bSi^{-1}-\bSi^{-1}\bSi_{(a)}\bSi^{-1}\bSi_{(b)}\bSi^{-1}+\bSi^{-1}\bSi_{(ab)}\bSi^{-1}$, which yields that $\tr(\W_a\bSi_{(b)})=\tr(\bSi^{-1}\bSi_{(a)}\bSi^{-1}\bSi_{(b)})=(\A)_{ab}$ and $(\B)_{ab}=\tr(\W_a\bSi\W_b\bSi)=\tr(\bSi^{-1}\bSi_{(a)}\bSi^{-1}\bSi_{(b)})=(\A)_{ab}$.
Thus, $\A^{-1}\B\A^{-1}=\A^{-1}$ and the covariance matrix of $\bpsih$ is $2\A^{-1}+O(N^{-3/2})$.
Moreover, note that 
\begin{align*}
(\K_a)_{bc}=& \tr(\W_{a(b)}\bSi\W_c\bSi)= -2\tr(\bSi^{-1}\bSi_{(a)}\bSi^{-1}\bSi_{(b)}\bSi^{-1}\bSi_{(c)})+\tr(\bSi^{-1}\bSi_{(ab)}\bSi^{-1}\bSi_{(c)}),\\
(\H_a)_{bc}=& \tr(\W_{a(b)}\bSi_{(c)})=-2\tr(\bSi^{-1}\bSi_{(a)}\bSi^{-1}\bSi_{(b)}\bSi^{-1}\bSi_{(c)})+\tr(\bSi^{-1}\bSi_{(ab)}\bSi^{-1}\bSi_{(c)}),
\end{align*}
which shows that $\W_a^{\rm REML}$ satisfies (\ref{eqn:cond}).

Case of $\W_a=(\bSi^{-1}\bSi_{(a)}+\bSi_{(a)}\bSi^{-1})/2$.
From (\ref{eqn:AB}), it follows that $(\A)_{ab}=\tr(\bSi^{-1}\bSi_{(a)}\bSi_{(b)})$ and $(\B)_{ab}=\{\tr(\bSi_{(a)}\bSi_{(b)})+\tr(\bSi^{-1}\bSi_{(a)}\bSi\bSi_{(b)})\}/2$.
The asymptotic covariance matrix of $\bpsih$ is $2\A^{-1}\B\A^{-1}$, and the bias is derived from (\ref{eqn:bias}).

Case of $\W_a=\bSi_{(a)}$.
Straightforward calculation shows that $(\A)_{ab} = \tr(\bSi_{(a)}\bSi_{(b)})$ and $(\B)_{ab}=\tr(\bSi_{(a)}\bSi\bSi_{(b)}\bSi)$.
The asymptotic covariance matrix of $\bpsih$ is $2 \A^{-1}\B\A^{-1}+O(N^{-3/2})$.
Moreover, since $W_{a(b)}=0$, the condition (\ref{eqn:cond}) holds.

\section{Summary of Estimation Methods in Specific Models}
Here we provide specific forms of the REML-type, FH-type, and their OLS-based estimators, the PR-type estimator and the Prasad-Rao estimator in the Fay-Herriot model and the nested error regression model.

\subsection{Fay-Herriot model}
The marginal distribution of $\y=(y_1, \ldots, y_m)^\top$ in the Fay-Herriot model has $E[\y]=\X\bbe$ and $\Cov(\y)=\bSi=\psi_1\I_m + \D$, where $p$ is a dimension of $\bbe$ and $\D=\diag(D_1, \ldots, D_m)$.

REML $\psih_1^{\rm RE}$ corresponds to $\W_1^{\rm RE}=\bSi^{-2}$ and $\bbeh=\bbeh^{\rm G}$ and the estimating equation is $(\y-\X\bbeh^{\rm G})^\top\bSi^{-2}(\y-\X\bbeh^{\rm G})=\tr(\P)$ for $\P=\bSi^{-1}-\bSi^{-1}\X(\X^\top\bSi^{-1}\X)^{-1}\X^\top\bSi^{-1}$. 

OLS-based REML $\psih_1^{\rm ORM}$ corresponds to $\W_1^{\rm RE}=\bSi^{-2}$ and $\bbeh=\bbeh^{\rm O}$ and the estimating equation is $(\y-\X\bbeh^{\rm O})^\top\bSi^{-2}(\y-\X\bbeh^{\rm O})=\tr(\bPt\bSi^{-2}\bPt\bSi)$ for $\bPt=\I-\X(\X^\top \X)^{-1}\X^\top$.

Fay-Herriot estimator $\psih_1^{\rm FH}$ corresponds to $\W_1^{\rm FH}=\bSi^{-1}$ and $\bbeh=\bbeh^{\rm G}$ and the estimating equation is $(\y-\X\bbeh^{\rm G})^\top\bSi^{-1}(\y-\X\bbeh^{\rm G})=m-p$.

OLS-based FH estimator $\psih_1^{\rm OFH}$ corresponds to $\W_1^{\rm FH}=\bSi^{-1}$ and $\bbeh=\bbeh^{\rm O}$ and the estimating equation is $(\y-\X\bbeh^{\rm O})^\top\bSi^{-1}(\y-\X\bbeh^{\rm O})=m-2p+\tr\{(\X^\top\X)^{-1}\X^\top\bSi\X(\X^\top\X)^{-1}\X^\top\bSi^{-1}\X\}$.

Prasad-Rao estimator $\psih_1^{\rm PR}$ corresponds to $\W_1^{\rm Q}=\I$ and $\bbeh=\bbeh^{\rm O}$ and it is given by $\psih_1^{\rm PR}=[\y^\top \bPt \y - \tr(\D)+\tr\{(\X^\top\X)^{-1}\X^\top\D\X\}]/(m-p)$.

The asymptotic variances and second-order biases can be provided from Proposition \ref{prp:FH} as follows:
REML $\psih_1^{\rm RE}$ and OLS-based REML $\psih_1^{\rm ORM}$ have the same asymptotic variance and the second-order bias
\begin{align*}
\Var(\psih_1^{\rm RE})\approx&{2 \over \tr(\bSi^{-2})} + {K_e\tr(\bSi^{-4}\D^2)+\tau^4K_v\tr(\bSi^{-4}) \over \{\tr(\bSi^{-2})\}^2},\\
{\rm Bias}(\psih_1^{\rm RE})\approx&-2{K_e \tr(\bSi^{-5}\D^2)+\tau^4K_v\tr(\bSi^{-5})\over \{\tr(\bSi^{-2})\}^2} + 2{\tr(\bSi^{-3})\{K_e \tr(\bSi^{-4}\D^2)+\tau^4K_v\tr(\bSi^{-4})\}\over \{\tr(\bSi^{-2})\}^3}.
\end{align*}
Fay-Herriot estimator $\psih_1^{\rm FH}$ and OLS-based FH estimator $\psih_1^{\rm OFH}$ have the same asymptotic variance and the second-order bias
\begin{align*}
\Var(\psih_1^{\rm FH})\approx&{2m \over \{\tr(\bSi^{-1})\}^2 } + {K_e\tr(\bSi^{-2}\D^2)+\tau^4K_v\tr(\bSi^{-2}) \over \{\tr(\bSi^{-1})\}^2},\\
{\rm Bias}(\psih_1^{\rm FH})\approx&
2{m\tr(\bSi^{-2})-\{\tr(\bSi^{-1})\}^2 \over \{\tr(\bSi^{-1})\}^3}\\
&-{K_e \tr(\bSi^{-3}\D^2)+\tau^4K_v\tr(\bSi^{-3})\over \{\tr(\bSi^{-1})\}^2} + {\tr(\bSi^{-2})\{K_e \tr(\bSi^{-2}\D^2)+\tau^4K_v\tr(\bSi^{-2})\}\over \{\tr(\bSi^{-1})\}^3}.
\end{align*}
Prasad-Rao estimator $\psih_1^{\rm PR}$ is second-order unbiased and has the asymptotic variance $\Var(\psih_1^{\rm PR})\approx\{2\tr(\bSi^2)+K_e\tr(\D^2)+m\tau^4K_v\}/m^2$.

\subsection{Nested error regression model}

The NER model is written as $\y_i=\X_i\bbe+\j_{n_i}v_i + \bep_i$ for $i=1, \ldots, m$, where $\y_i$, $\bbe$ and $\bep_i$ are $n_i$, $p$ and $n_i$ dimensional vectors, $\X_i$ is an $n_i\times p$ matrix, $v_i$ is scalar and $\j_{n_i}=(1, \ldots, 1)^\top\in \Re^{n_i}$.
Here, $v_i$ and $\bep_i$ are independent random variables such that $E[v_i]=0$, $\Var(v_i)=\psi_1$, $E[\bep_i]=\zero$ and $\Cov(\bep_i)=\psi_2\I_{n_i}$.
Let $\y=(\y_1^\top, \ldots, \y_m^\top)^\top$, $\X=(\X_1^\top, \ldots, \X_m^\top)^\top$, $N=\sum_{i=1}^m n_i$ and $\G=\text{block diag}(\J_{n_1}, \ldots, \J_{n_m})$ for $\J_{n_i}=\j_{n_i}\j_{n_i}^\top$.
Also, let $\bSi=\text{block diag}(\bSi_1, \ldots, \bSi_m)$ for $\bSi_i=\psi_1\J_{n_i}+\psi_2\I_{n_i}$.
Then, $\bSi=\psi_1\G+\psi_2\I_N$, $\bSi_{(1)}=\G$ and $\bSi_{(2)}=\I_N$.

REML $\psih_1^{\rm RE}$ and $\psih_1^{\rm RE}$ correspond to $\W_1^{\rm RE}=\bSi^{-1}\G\bSi^{-1}$, $\W_2^{\rm RE}=\bSi^{-2}$ and $\bbeh=\bbeh^{\rm G}$, and the estimating equations are $(\y-\X\bbeh^{\rm G})^\top\bSi^{-1}\G\bSi^{-1}(\y-\X\bbeh^{\rm G})=\tr(\P\G)$ and $(\y-\X\bbeh^{\rm G})^\top\bSi^{-2}(\y-\X\bbeh^{\rm G})=\tr(\P)$.

OLS-based REML $\psih_1^{\rm ORM}$ and $\psih_2^{\rm ORM}$ correspond to $\W_1^{\rm RE}=\bSi^{-1}\G\bSi^{-1}$, $\W_2^{\rm RE}=\bSi^{-2}$ and $\bbeh=\bbeh^{\rm O}$, and the estimating equations are $(\y-\X\bbeh^{\rm O})^\top\bSi^{-1}\G\bSi^{-1}(\y-\X\bbeh^{\rm O})=\tr(\bPt\bSi\bPt \bSi^{-1}\G\bSi^{-1})$ and $(\y-\X\bbeh^{\rm O})^\top\bSi^{-2}(\y-\X\bbeh^{\rm O})=\tr(\bPt\bSi\bPt \bSi^{-2})$.

FH-type estimators $\psih_1^{\rm FH}$ and $\psih_2^{\rm FH}$ correspond to $\W_1^{\rm FH}=(\bSi^{-1}\G+\G\bSi^{-1})/2$, $\W_2^{\rm FH}=\bSi^{-1}$ and $\bbeh=\bbeh^{\rm G}$, and the estimating equations are 
\begin{align*}
&\sum_{i=1}^m{n_i^2(\yo_i-\xb_i^\top\bbeh^{\rm G})^2\over n_i\psi_1+\psi_2}= N - \sum_{i=1}^m {n_i^2\xb_i^\top(\X^\top\bSi^{-1}\X)^{-1}\xb_i\over n_i\psi_1+\psi_2},\\
&\psi_2={1\over N-p}\sum_{i=1}^m\sum_{j=1}^{n_i}(y_{ij}-\x_{ij}^\top\bbeh^{\rm G})^2-{1\over N-p}\sum_{i=1}^m{n_i^2\psi_1\over n_i\psi_1+\psi_2}(\yo_i-\xb_i^\top\bbeh^{\rm G})^2.
\end{align*}

OLS-based FH estimators $\psih_1^{\rm OFH}$ and $\psih_2^{\rm OFH}$ correspond to $\W_1^{\rm FH}=(\bSi^{-1}\G+\G\bSi^{-1})/2$, $\W_2^{\rm FH}=\bSi^{-1}$ and $\bbeh=\bbeh^{\rm O}$, and the estimating equations are 
\begin{align*}
&\sum_{i=1}^m{n_i^2(\yo_i-\xb_i^\top\bbeh^{\rm O})^2\over n_i\psi_1+\psi_2}= N - 2\sum_{i=1}^m n_i^2\xb_i^\top(\X^\top\X)^{-1}\xb_i+\sum_{i=1}^m {n_i^2\xb_i^\top(\X^\top\X)^{-1}\X^\top\bSi\X(\X^\top\X)^{-1}\xb_i\over n_i\psi_1+\psi_2},\\
&\sum_{i=1}^m\sum_{j=1}^{n_i}(y_{ij}-\x_{ij}^\top\bbeh^{\rm O})^2-\sum_{i=1}^m{n_i^2\psi_1\over n_i\psi_1+\psi_2}(\yo_i-\xb_i^\top\bbeh^{\rm O})^2 =\tr(\bPt\bSi\bPt\bSi^{-1}).
\end{align*}

PR-type estimators $\psih_1^{\rm Q}$ and $\psih_2^{\rm Q}$ correspond to $\W_1^{\rm Q}=\G$, $\W_2^{\rm Q}=\I$ and $\bbeh=\bbeh^{\rm O}$, and the estimators are 
$\psih_1^{\rm Q} = \{\sum_{i=1}^m n_i^2 (\yo_i-\xb_i^\top\bbeh^{\rm O})^2-\psih_2\tr(\bPt \G)\}/\tr(\bPt \G)^2$ and 
$$
\psih_2^{\rm Q} ={ \sum_{i=1}^m\sum_{j=1}^{n_i}(y_{ij}-\x_{ij}^\top\bbeh^{\rm O})^2-
[\tr(\bPt\G)/\tr\{(\bPt\G)^2\}]\sum_{i=1}^m n_i^2 (\yo_i-\xb_i^\top\bbeh^{\rm O})^2
\over N-p-\{\tr(\bPt\G)\}^2\tr\{(\bPt\G)^2\}}.
$$

Prasad-Rao estimators are $\psih_1^{\rm PR}=\{\y^\top\bPt\y-(N-p)\psih_2\}/\{N-\sum_{i=1}^mn_i^2\xb_i^\top(\X^\top\X)^{-1}\xb_i\}$ and $\psih_2^{\rm PR}=\{ \y^\top\{\E-\E\X(\X^\top\E\X)^{-1}\X^\top\E\}\y\}/(N-k-p)$, where $\E=\text{block diag}(\I_{n_1}-n_1^{-1}\J_{n_1}, \ldots, \I_{n_m}-n_m^{-1}\J_{n_m})$.

Hereafter we assume that $K_e=K_v=0$ for simplicity.
Note that $\bSi\G=\G\bSi$, $\psi_1\G=\bSi-\psi_2\I_N$, $\psi_2\bSi^{-1}=\I_N-\psi_1\text{block diag}(\ga_1\J_{n_1}, \ldots, \ga_m\J_{n_m})$, $\psi_2^2\bSi^{-2}=\I_N-\psi_1\text{block diag}((1+\psi_2\ga_1)\ga_1\J_{n_1}, \ldots, (1+\psi_2\ga_m)\ga_m\J_{n_m})$ for $\ga_i=1/(\psi_2+n_i\psi_1)$. 
Then the asymptotic variances and second-order biases can be provided from Proposition \ref{prp:NER} as follows:
REML $\bpsih^{\rm RE}$ and OLS-based REML $\bpsih^{\rm ORM}$ are second-order unbiased and have the same asymptotic variance 
\begin{align*}
\Cov(\bpsih^{\rm RE})\approx&
2 \begin{pmatrix}\tr\{(\bSi^{-1}\G)^2\} & \tr(\bSi^{-2}\G)\\ \tr(\bSi^{-2}\G) & \tr(\bSi^{-2})\end{pmatrix}^{-1}
=2 \begin{pmatrix}\sum_{i=1}^mn_i^2\ga_i^2 & \sum_{i=1}^mn_i\ga_i^2\\ \sum_{i=1}^mn_i\ga_i^2 & (N-m)/\psi_2^2+\sum_{i=1}^m\ga_i^2\end{pmatrix}^{-1},
\end{align*}
Which was given in Datta and Lahiri (2000).

Fay-Herriot estimator $\bpsih^{\rm FH}$ and OLS-based FH estimator $\bpsih^{\rm OFH}$ have the same asymptotic covariance matrix $\Cov(\bpsih^{\rm FH})\approx2\A_{\rm FH}^{-1}\B_{\rm FH}\A_{\rm FH}^{-1}$, where
\begin{align*}
\A_{\rm FH}=&
\begin{pmatrix}\tr(\bSi^{-1}\G^2) & \tr(\bSi^{-1}\G)\\ \tr(\bSi^{-1}\G) & \tr(\bSi^{-1})\end{pmatrix}
=\begin{pmatrix}\sum_{i=1}^mn_i^2\ga_i & \sum_{i=1}^mn_i\ga_i\\ \sum_{i=1}^mn_i\ga_i & (N-m)/\psi_2+\sum_{i=1}^m\ga_i\end{pmatrix},\\
\B_{\rm FH}=&
\begin{pmatrix}\tr(\bSi^{-1}\G\bSi\G+\G^2) & N\\ N & N\end{pmatrix}
=\begin{pmatrix}\sum_{i=1}^mn_i^2 & N\\ N & N\end{pmatrix},
\end{align*}
and the same second-order bias
$$
{\bf Bias}(\bpsih^{\rm FH})\approx
2\A_{\rm FH}^{-1} \begin{pmatrix} \tr(\K_1\A_{\rm FH}^{-1})-\tr(\H_1\A_{\rm FH}^{-1}\B_{\rm FH}\A_{\rm FH}^{-1})\\ \tr(\K_2\A_{\rm FH}^{-1})-\tr(\H_2\A_{\rm FH}^{-1}\B_{\rm FH}\A_{\rm FH}^{-1})\end{pmatrix},
$$
where
\begin{align*}
\K_1=&-\begin{pmatrix} \tr(\bSi^{-1}\G^3)& \tr(\bSi^{-1}\G^2)\\ \tr(\bSi^{-1}\G^2)& \tr(\bSi^{-1}\G)\end{pmatrix}
=-\begin{pmatrix} \sum_i n_i^3\ga_i & \sum_i n_i^2\ga_i \\ \sum_i n_i^2\ga_i & \sum_i n_i\ga_i\end{pmatrix},\\
\K_2=&-\begin{pmatrix} \tr(\bSi^{-1}\G^2)& \tr(\bSi^{-1}\G)\\ \tr(\bSi^{-1}\G)& \tr(\bSi^{-1})\end{pmatrix}
=-\begin{pmatrix} \sum_i n_i^2\ga_i & \sum_i n_i\ga_i \\ \sum_i n_i\ga_i & (N-m)/\psi_2+\sum_i \ga_i\end{pmatrix},\\
\H_1=&-\begin{pmatrix} \tr(\bSi^{-1}\G\bSi^{-1}\G^2)& \tr(\bSi^{-1}\G\bSi^{-1}\G)\\ \tr(\bSi^{-2}\G^2)& \tr(\bSi^{-2}\G)\end{pmatrix}
=-\begin{pmatrix} \sum_i n_i^3\ga_i^2 & \sum_i n_i^2\ga_i^2 \\ \sum_i n_i^2\ga_i^2 & \sum_i n_i\ga_i^2\end{pmatrix},\\
\H_2=&-\begin{pmatrix} \tr(\bSi^{-1}\G\bSi^{-1}\G)& \tr(\bSi^{-2}\G)\\ \tr(\bSi^{-2}\G)& \tr(\bSi^{-2})\end{pmatrix}
=-\begin{pmatrix} \sum_i n_i^2\ga_i^2 & \sum_i n_i\ga_i^2 \\ \sum_i n_i\ga_i^2 & (N-m)/\psi_2+\sum_i \ga_i^2\end{pmatrix}.
\end{align*}
PR-type estimator $\bpsih^{\rm Q}$ is second-order unbiased and has the same asymptotic covariance matrix $\Cov(\bpsih^{\rm Q})\approx2\A_{\rm Q}^{-1}\B_{\rm Q}\A_{\rm Q}^{-1}$, where
\begin{align*}
\A_{\rm Q}=&
\begin{pmatrix}\tr(\G^2) & \tr(\G)\\ \tr(\G) & \tr(\I_N)\end{pmatrix}
=\begin{pmatrix}\sum_{i=1}^mn_i^2 & N \\ N & N \end{pmatrix},\\
\B_{\rm Q}=&
\begin{pmatrix}\tr(\bSi^2\G^2) & \tr(\bSi^2\G)\\ \tr(\bSi^2\G) & \tr(\bSi^2)\end{pmatrix}
=\begin{pmatrix}\sum_{i=1}^mn_i^2/\ga_i^2 & \sum_{i=1}^mn_i/\ga_i^2\\ \sum_{i=1}^mn_i/\ga_i^2 & (N-m)\psi_2^2+\sum_{i=1}^m1/\ga_i^2\end{pmatrix},
\end{align*}

\end{document}